\documentclass[utf8]{frontiersSCNS} 
\usepackage{url,lineno, microtype, subcaption}
\usepackage[onehalfspacing]{setspace}
\usepackage{xcolor}
\usepackage{caption}
\definecolor{bg}{rgb}{.96, .96, .96}

\def\firstAuthorLast{Oliveira {et~al.}} 
\def\Authors{Denny M. Oliveira\,$^{1,2,*}$}

\def\Address{$^{1}$Goddard Planetary Heliophysics Institute, University of Maryland, 
				   Baltimore County, Baltimore, MD, United States
			 $^{2}$Geospace Physics Laboratory, NASA Goddard Space Flight Center, Greenbelt, MD, United States
			 }

\def\corrAuthor{Denny M. Oliveira}
\def\corrEmail{denny@umbc.edu}

\def\dbdt{d$B$/d$t$}
\def\thxn{$\theta_{x_n}$}
\def\thbn{$\theta_{B_n}$}
\def\phiyn{$\varphi_{y_n}$}
\def\vs{$v_s$}
\def\Ma{$M_A$}
\def\Ms{$M_s$}
\def\Xn{$X_{N}$}
\def\Xdp{$X_{dp}$}
\def\Xb{$X_{B}$}

\usepackage[breaklinks, colorlinks = true,
            linkcolor = SkyBlue,
            urlcolor = SkyBlue,
            citecolor = SkyBlue,
            anchorcolor = white,
            breaklinks]{hyperref}

\begin{document}
\onecolumn
\firstpage{1}

\title[Shock data base]{Interplanetary Shock Data Base} 

\author[\firstAuthorLast ]{\Authors} 
\address{} 
\correspondance{} 

\extraAuth{}

\maketitle

\section{Introduction}
	
	Interplanetary (IP) shocks are frequently observed in the solar wind \citep{Burlaga1971a,Stone1985}. Many levels of different kinds of geomagnetic activity may follow the impact of IP shocks on the Earth's magnetosphere. Such effects are seen everywhere in the magnetosphere-ionosphere system, including radiation belt dynamics, magnetic field in geosynchronous orbit, field-aligned currents, ionospheric disturbances, satellite orbital drag, ground magnetometers, geomagnetically induced currents (GICs), and others \citep[e.g., ][]{Echer2005b,Tsurutani2011a,Khazanov2016,Oliveira2017d,Oliveira2019b,Abda2020,Smith2020a,Bhaskar2021}. The study of IP shocks is important for space weather purposes because shock impacts occur more frequently than geomagnetic storms and correlate well with solar activity \citep{Oh2007,Kilpua2015a,Echer2023}. Therefore, keeping an updated and accurate IP shock data base is of primary importance to the scientific community. \par

	Many IP shock parameters control shock geoeffectiveness, such as shock speeds, Mach numbers, and compression ratios \citep{Craven1986,Kabin2001,Goncharov2014}. Additionally, the shock impact angle, the angle the shock normal vector performs with the Sun-Earth line, has been shown to be a significant factor that controls shock geoeffectiveness \citep{Oliveira2018a,Oliveira2023b}. Many works have demonstrated with simulations and observations that, in general, the more frontal and the faster the shock, the higher the subsequent geomagnetic activity observed from the geospace to the ground \citep[e.g.,][]{Takeuchi2002b,Guo2005,Wang2006a,Oliveira2014b,Samsonov2015,Oliveira2015a,Oliveira2016a,Selvakumaran2017,Oliveira2018b,Baker2019,Rudd2019,Shi2019b,Oliveira2020d,Xu2020a,Oliveira2021b}. \par

	The main goal of this short report is to release an expanded version of an IP shock data base that was published before \citep{Oliveira2015a,Oliveira2018b}. A major component of this new shock data base is a revision of the methodology used to calculate shock impact angles and speeds with respect to past versions of this list. Additionally, more shock and solar wind parameters before and after shock impacts and geomagnetic activity information were included in the list. \par

	This article is organized as follows. Section 2 discusses the methodology used for the computation of shock properties, including the data used and shock normal calculation methods. A shock example is shown in Section 3. Section 4 presents the IP shock data base and its components. Finally, section 5 brings a few suggestions for future usage of this shock list, with focus on the role of shock impact angles in controlling the subsequent shock geoeffectiveness.

\section{Methods}

	\subsection{Solar wind plasma and IMF data}

		Properties and normal vector orientations of IP shocks are computed with the use of solar wind plasma and interplanetary magnetic field (IMF) data collected by solar wind monitors upstream of the Earth at the Lagrangian point L1. The time coverage of this shock list ranges from January 1995 to May 2023. Wind plasma data are collected by the Solar Wind Experiment instrument with resolution of 92 s \citep{Ogilvie1995}, and Wind magnetic field data are collected by the Magnetic Field Investigation instrument with resolution of 3 s \citep{Lepping1995}. ACE (Advanced Composition Explorer) collects solar wind data (resolution 64 s) with the Solar Wind Electron, Proton and Alpha Monitor instrument \citep{McComas1998}, and magnetic field data (resolution 16 s) with the MAG magnetometer instrument \citep{Smith1998}. All the data used for computations is represented in geocentric solar ecliptic (GSE) coordinates.

	\vspace{1cm}
	\subsection{SuperMAG ground magnetometer and sunspot number data}

		Supporting geomagnetic index data are provided by the SuperMAG initiative \citep{Gjerloev2009}. SuperMAG computes geomagnetic indices using larger numbers of magnetometers in comparison to traditional IAGA (International Association of Geomagnetism and Aeronomy) indices \citep{Davis1966,Rostoker1972}. The SuperMAG ring current index, SMR, is explained by \cite{Newell2012}, and the SuperMAG auroral indices SME, SMU, and SML are documented in \cite{Newell2011a}. SMU is the upper envelope index, SML is the lower envelope index, and SME = SMU -- SML. All SuperMAG index data have resolution of 1 minute. 


		Another supporting data set, with daily sunspot number observations, is provided by Sunspot Index and Long-term Solar Observations, Royal Observatory of Belgium, Brussels \citep{Clette2016b}. The sunspot number data base used in this report has been corrected and recalibrated according to the methods explained by \cite{Clette2016b}. 

		\vspace{1cm}

	\subsection{Computations of shock normals and shock-related parameters}

		For the purpose of shock normal computations at 1 AU, shock fronts are assumed to be planar structures larger than the Earth's magnetospheric system \citep{Russell1983,Russell2000b}. Then, IP shock normal vectors can be computed if data from at least one spacecraft is available \citep{Russell1983,Aguilar-Rodriguez2010,Trotta2023}. Generally, shocks driven by CMEs (coronal mass ejections) have their shock normals with small deviation with respect to the Sun-Earth line, whereas shocks driven by CIRs (corotating interacting regions) have their shock normals with large deviations from the Sun-Earth line \citep{Kilpua2015a,Oliveira2018a}. Such shock inclinations occur because CMEs tend to travel radially in the solar wind, while CIRs tend to follow the Parker spiral when slow speed streams are compressed by fast speed streams \citep{Pizzo1991,Tsurutani2006a,Cameron2019a}. An animation showing the different inclinations of a CME-driven shock and a CIR- driven shock can be accessed here: \url{https://dennyoliveira.weebly.com/phd.html}. \par

		There are three different ways commonly used to compute shock normal orientations. They use magnetic field data only, solar wind velocity data only, and a combination of magnetic field and solar wind velocity data. Such methods are, respectively, named magnetic coplanarity \citep[MC,][]{Colburn1966}, velocity coplanarity \citep[VC,][]{Abraham-Shrauner1972}, and three mixed  data methods \citep[MX1, MX2, MX3,][]{Schwartz1998}. The equations used are listed below:

		\begin{eqnarray}
			\vec{n}_{MC}  &=& \pm\frac{(\vec{B}_2 \times \vec{B}_1) \times (\vec{B}_2 - \vec{B_1})}{|(\vec{B}_2 \times \vec{B}_1) \times (\vec{B}_2 - \vec{B_1})|} \\
			\vec{n}_{MX1} &=& \pm\frac{\vec{B}_1 \times (\vec{V}_2 - \vec{V}_1) \times (\vec{B}_2 - \vec{B_1})}{|\vec{B}_1 \times (\vec{V}_2 - \vec{V}_1) \times (\vec{B}_2 - \vec{B_1})|} \\
			\vec{n}_{MX2} &=& \pm\frac{\vec{B}_2 \times (\vec{V}_2 - \vec{V}_1) \times (\vec{B}_2 - \vec{B_1})}{|\vec{B}_2 \times (\vec{V}_2 - \vec{V}_1) \times (\vec{B}_2 - \vec{B_1})|} \\
			\vec{n}_{MX3} &=& \pm\frac{(\vec{B}_2 - \vec{B}_1) \times (\vec{V}_2 - \vec{V}_1) \times (\vec{B}_2 - \vec{B_1})}{|(\vec{B}_2 - \vec{B}_1) \times (\vec{V}_2 - \vec{V}_1) \times (\vec{B}_2 - \vec{B_1})|} \\
			\vec{n}_{VC} &=& \pm\frac{\vec{V}_2 - \vec{V}_1}{|\vec{V}_2 - \vec{V}_1|}
		\end{eqnarray}

		In these equations, $\vec{B}$ is the magnetic field vector, and $\vec{V}$ is the solar wind velocity vector. Indices 1 and 2 represent the upstream (non-shocked) region, and downstream (shocked) region, behind of and ahead the shock, respectively. The sign of each vector $\vec{n}$ is arbitrary and can be chosen to indicate whether the normal vector points toward the downstrean direction (+) or upstream direction (--) \citep{Schwartz1998}. \par

		Equations (1-5) provide a three-dimensional normal vector $\vec{n} = (n_x, n_y, n_z)$ in Cartesian coordinates, from which three angles can be extracted:
		\begin{eqnarray}
			\theta_{x_n} &=& \cos^{-1}(n_x)\,, \\
			\varphi_{y_n} &=& \tan^{-1}\left( \frac{n_z}{n_y} \right)\,, \\
			\theta_{B_n} &=& \frac{\vec{n} \cdot \vec{B}_1}{|\vec{B}_1|}\,,
		\end{eqnarray}
		where \thxn{} is named the shock impact angle, the angle the shock normal vector performs with the Sun-Earth line,  \phiyn{} is the shock clock angle in the yz plane perpendicular to the Sun-Earth line (both angles in the satellite or Earth reference frame), and \thbn{} is the angle between the upstream magnetic field vector and the shock normal vector (in the shock reference frame). \par

		\begin{figure*}
			\centering
			\includegraphics[width = 17cm]{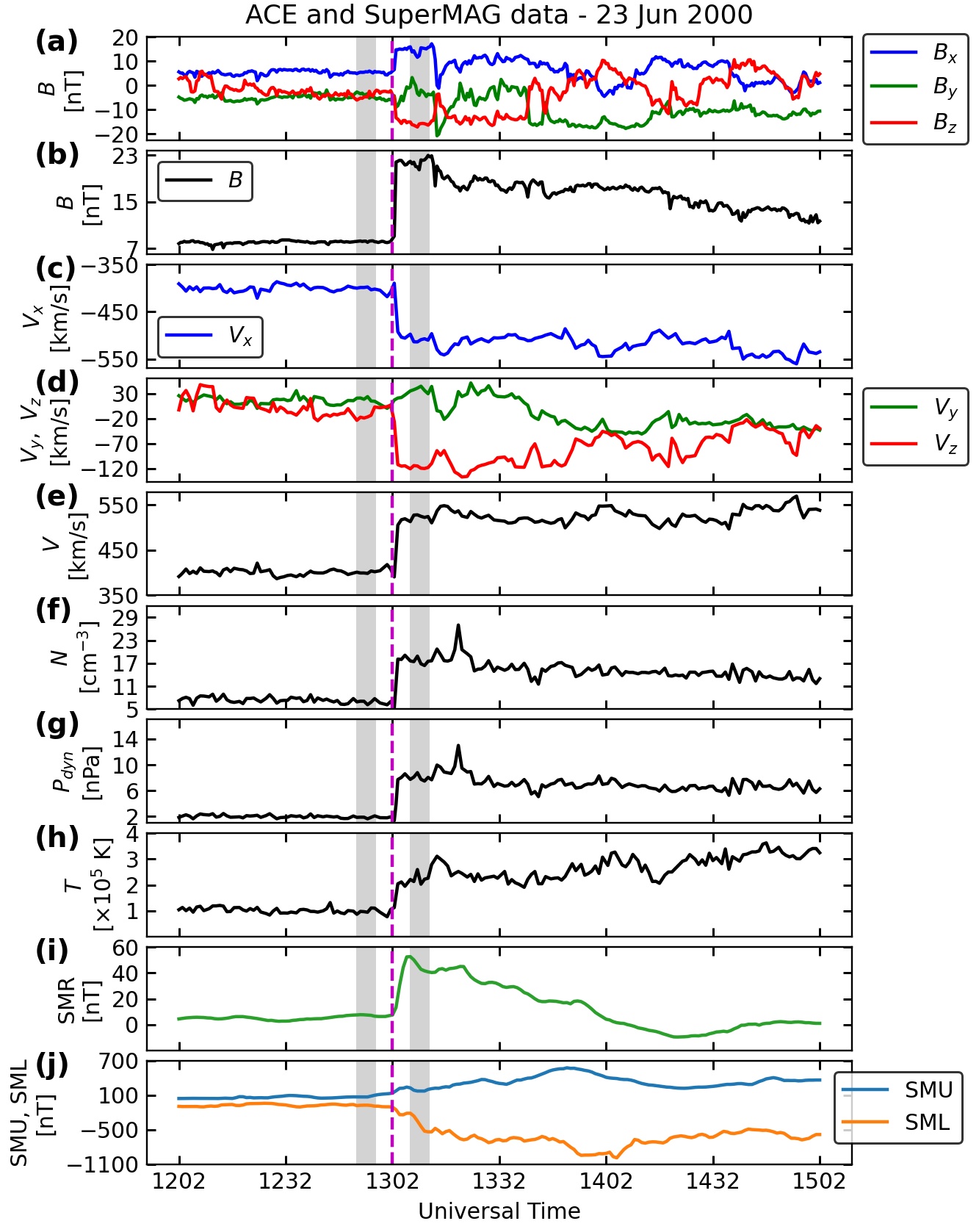}
			\caption{Interplanetary shock observed by ACE on 23 June 2000 and the subsequent geomagnetic activity represented by SuperMAG data.}
			\label{shock_142}
		\end{figure*}

		Since the plasma mass flux must be conserved along the shock normal, $\rho_1u_{n1} = \rho_2u_{n2}$, with $u_{n1,n2} = v_s - \vec{V}_{1,2}\cdot\vec{n}$, the shock speed is computed as follows:
		\begin{equation}
			v_s = \vec{n}\cdot\left(  \frac{\vec{V}_2\rho_2 - \vec{V}_1\rho_1}{\rho_2 - \rho_1} \right)
		\end{equation}

		Other useful shock velocities are represented by:
		
		\begin{align}
			c_s &= \sqrt{\frac{\gamma P_1}{\rho_1}} & \hspace{3cm}\mbox{sound speed,} \\
			v_A &= \frac{|\vec{B}_1|}{\sqrt{\mu_0 \rho_1}} & \hspace{3cm}\mbox{Alfv\'en speed,} \\
			v_{ms} &= \frac{1}{2}\sqrt{v_A^2 + c_s^2 \pm \sqrt{(v_A^2 + c_s^2)^2 - 4v_A^2c_s^2\cos^2(\theta_{B_n})}} & \hspace{3cm}\mbox{magnetosonic speed,}
			\label{mag_speed}
		\end{align}
		where $\gamma = 5/3$ is the ratio of the solar wind heat capacity with constant pressure to the heat capacity with constant volume; $P_1$ is the upstream solar wind thermal pressure; $\rho_1$ is the upstream solar wind density; and $\mu_0 = 4\pi\times10^{-7}$ N/A$^2$ is the magnetic vacuum permeability. The positive solution of equation \ref{mag_speed} gives the fast magnetosonic speed, whereas the negative solution yields the slow magnetosonic speed \citep{Jeffrey1964,Priest1981,Boyd2003}. \par

		Shock strengths are usually represented by specific Mach numbers. With $u$ = $v_s - \vec{V}\cdot\vec{n}$ being the relative speed between the shock speed and the local solar wind velocity, the Mach numbers are represented by:
		\begin{align}
			& M_A = \frac{u}{v_A} & \hspace{2.25cm} & \mbox{Alfv\'enic Mach number,} \\
			& M_s = \frac{u}{v^f_{ms}} & \hspace{2.25cm} & \mbox{fast magnetosonic Mach number,}
		\end{align}
		where $v^f_{ms}$ is the fast magnetosonic speed. \par

		Finally, the strength of IP shocks can also be indicated by upstream and downstream solar wind plasma parameters and IMF. Such compression ratios are represented by:
		\begin{align}
			& X_n = \frac{n_2}{n_1} & \hspace{2.5cm} & \mbox{plasma number density compression ratio,} \\
			& X_{dp} = \frac{\rho_2V_2^2}{\rho_1V_1^2} & \hspace{2.5cm} & \mbox{dynamic pressure compression ratio,} \\
			& X_B = \frac{|\vec{B}_2|}{|\vec{B}_1|} & \hspace{2.5cm} & \mbox{magnetic field compression ratio.} 
		\end{align}

	\section{The IP shock of 23 June 2000 as an example}

		Figure \ref{shock_142}, first published by \cite{Oliveira2015a}, shows an IP shock event that occurred on 23 June 2000 observed by ACE at 1226 UT upstream of the Earth at (x, y, z) = (239.9, 36.7, --0.7) $R_E$, where $R_E$ is the Earth's radius = 6371.1 km. Solar wind plasma and IMF data are depicted in the figure, along with SuperMAG geomagnetic index data. From top to bottom, the plot shows three components of the IMF (a); IMF magnitude (b); x component of solar wind velocity (c), y and z components of solar wind velocity (d); solar wind velocity magnitude (e); solar wind particle number density (f); solar wind dynamic pressure $P_{dyn}=\rho V^2$ (g); solar wind thermal temperature (h); SMR index (i); and SMU/SML indices (j). The vertical dashed magenta lines indicate the time of shock impact on the magnetosphere. The data were shifted to the magnetopause nose to match shock observations with the onsets in the ground geomagnetic indices. The highlighted grey areas correspond to the shock upstream region (left) 10 to 5 minutes before shock impact, and shock downstream region (right), 5 to 10 minutes after shock impact. Average values of these regions are used in equations (1-5) for the computation of the shock normal orientations with the five different methods. Most shocks have the time windows mentioned above, but a few events have different time windows. \par

		As discussed by \cite{Balogh1995} and \cite{Trotta2022}, the length of the upstream and downstream windows around the shocks is an important factor in determining shock parameters. For example, \cite{Trotta2022} suggested a method to use windows with different lengths with short-length windows being located near the shock. This methodology provides statistical significance (including uncertainties) to the calculated shock parameters and reliability to the subsequent results. This approach will be applied to this shock list in a future work for further improvements of this shock data base. \par

		The data shown in the figure is processed before plotting and before computing shock parameters and normal orientations. First, bad data points, such as 1E+31 are replaced by nan (``not a number") values and subsequently linearly interpolated. Then, solar wind parameter data are interpolated, and IMF data are averaged to a uniform time cadence of 30 seconds. Differences between the non-interpolated and interpolated data are very small or nearly nonexistent around the shock onset. This process allows the time resolutions of both data sets to match to further perform computations that involve both data sets. The same technique was applied to all events in the shock data base. \par

		Positive step-like enhancements are seen in all solar wind plasma parameters and IMF. This is a clear signature of a fast forward IP shock \citep{Priest1981,Tsurutani2011a,Oliveira2017a}. More information on the analysis of this event including shock normal orientations will be provided in the next section.

\section{The IP shock data base}

	Previous versions of this current shock data base were published by \cite{Oliveira2015a} and \cite{Oliveira2018b}. A few sources were used to compile these previous shock lists: a shock catalog provided by the Harvard-Smithsonian Center for Astrophysics and compiled by Dr. J. C. Kasper for Wind (\url{http://www.cfa.harvard.edu/shocks/wi_data/}) and ACE (\url{ http://www.cfa.harvard.edu/shocks/ac_master_data/}); a shock list compiled by the ACE team (\url{at http://www-ssg.sr.unh.edu/mag/ace/ACElists/obs_list.html#shocks.}); and another list published by \cite{Wang2010d} with events from February 1998 to August 2008. New events were added by scanning solar wind and IMF data to detect shock events that satisfy the framework discussed in section 2.3. \par

	The IP shock data base consists of three files: (i) \texttt{full\_shock\_list\_2023.txt}, a text file with 603 events; (ii) \texttt{full\_shock\_params.cdf}, a cdf file with detailed information about each specific shock event; and (iii) \texttt{read\_shock.py}, a file that contains a short python routine to read information about a specific shock event. The SpacePy package \citep[\url{https://spacepy.github.io},][]{Morley2011,SpacePy2022} is required to extract shock information from the cdf file using \texttt{read\_shock.py}. \par

	\setcounter{figure}{0}
	\renewcommand{\figurename}{Listing}
	\renewcommand{\thefigure}{ \arabic{figure}}

	The Python routine to read the information of a specific shock event is




	\begin{figure}
		\includegraphics[width = \textwidth]{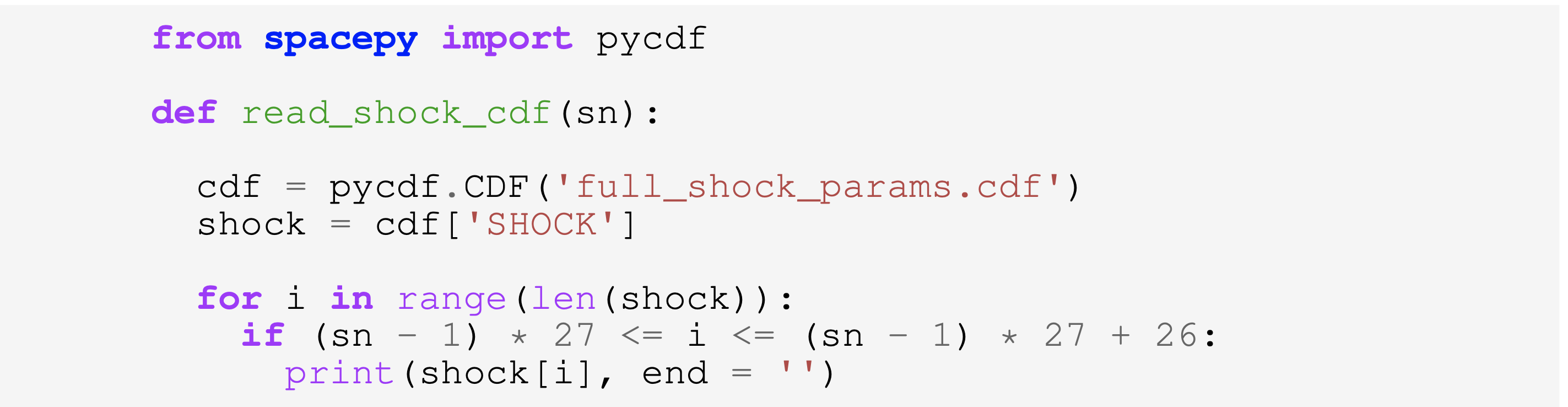}
		\caption{Python routine to extract information related to a specific shock event from the cdf file.}
	\end{figure}

	The input variable for the above routine is the shock number \texttt{sn}. The IP shock represented in Figure \ref{shock_142} is the event number 142 in the shock list. Therefore, \texttt{read\_shock.py} can be run as follows

	\begin{figure}
		\includegraphics[width = \textwidth]{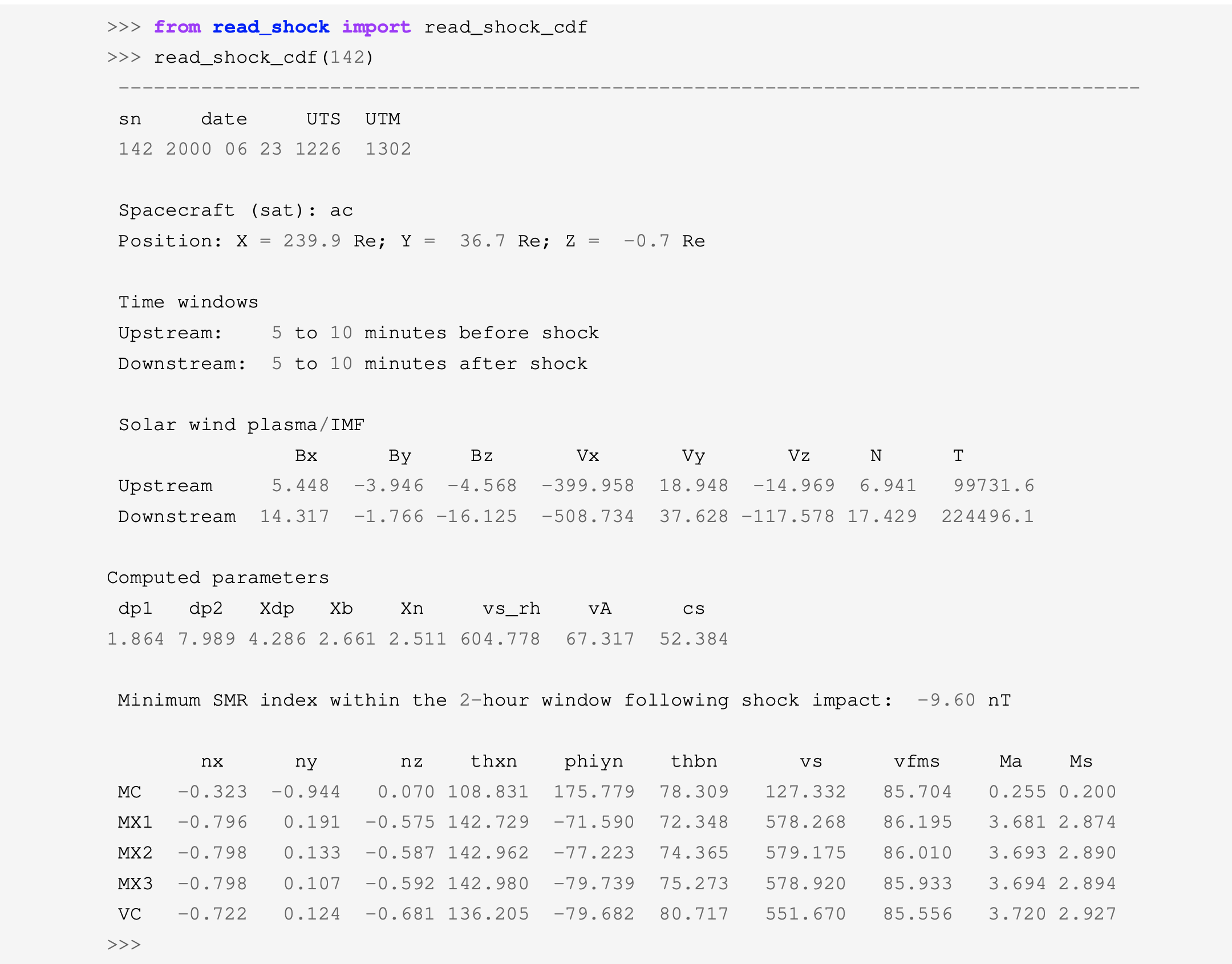}
		\caption{Example of how to run the Python routine \texttt{read\_shock.py} shown in Listing \ref{listing1} to extract information about a shock event in the list. The example shown in this listing is the event number 142, occurred on 23 June 2000 and observed by ACE (see Figure \ref{shock_142}). }
	\end{figure}

	The results for each shock are used to compose the shock list in the \texttt{full\_shock\_list\_2023.txt} file. The file brings a header with the names of the variables (Table \ref{table1}). The list also includes the position (in $R_E$) of the solar wind monitor (either Wind or ACE) whose data are used in the calculations, along with minimum SMR values occurring in a time window of two hours after shock impact. This time window was chosen because amplitudes of geomagnetic activity response usually occur $\sim$60 minutes after energy being released by the magnetotail \citep{Bargatze1985,Oliveira2015a,Oliveira2021b}. Such values can be used in studies that aim to use shock observations during non-storm times.

	\begin{table}
	\centering
	\begin{tabular}{r l c l}
		\hline
		1\hspace{1cm} & sn & \hspace{1.cm} & shock number; \\ 
		2\hspace{1cm} & YY & \hspace{1.cm} & year; \\ 
		3\hspace{1cm} & MM & \hspace{1.cm} & month; \\ 
		4\hspace{1cm} & DD & \hspace{1.cm} & day; \\ 
		5\hspace{1cm} & UTS & \hspace{1.cm} & UT of shock observation by solar wind monitor; \\ 
		6\hspace{1cm} & UTM & \hspace{1.cm} & UT of ground magnetic sudden impulse onset; \\ 
		7\hspace{1cm} & nx & \hspace{1.cm} & x component of shock normal vector; \\ 
		8\hspace{1cm} & ny & \hspace{1.cm} & y component of shock normal vector; \\ 
		9\hspace{1cm} & nz & \hspace{1.cm} & z component of shock normal vector; \\ 
		10\hspace{1cm} & thxn & \hspace{1.cm} & shock impact angle (degrees); \\
		11\hspace{1cm} & phiyn & \hspace{1.cm} & shock clock angle in the yz plane (degrees); \\
		12\hspace{1cm} & thbn & \hspace{1.cm} & shock obliquity angle (degrees); \\
		13\hspace{1cm} & vs & \hspace{1.cm} & shock speed (km/s); \\
		14\hspace{1cm} & cs & \hspace{1.cm} & sound speed (km/s); \\
		15\hspace{1cm} & vA & \hspace{1.cm} & Alfv\'en speed (km/s); \\
		16\hspace{1cm} & vfms & \hspace{1.cm} & fast magnetosonic speed (km/s); \\
		17\hspace{1cm} & Ma & \hspace{1.cm} & Alfv\'enic Mach number; \\
		18\hspace{1cm} & Ms & \hspace{1.cm} & magnetosonic Mach number; \\
		19\hspace{1cm} & dp1 & \hspace{1.cm} & upstream solar wind dynamic pressure (nPa); \\
		20\hspace{1cm} & dp2 & \hspace{1.cm} & downstream solar wind dynamic pressure (nPa); \\
		21\hspace{1cm} & Xn & \hspace{1.cm} & solar wind number density compression ratio; \\
		22\hspace{1cm} & Xdp & \hspace{1.cm} & solar wind dynamic pressure ratio (dp2/dp1); \\
		23\hspace{1cm} & Xb & \hspace{1.cm} & magnetic field compression ratio; \\
		24\hspace{1cm} & bz1 & \hspace{1.cm} & upstream z component of interplanetary magnetic field (nT); \\
		25\hspace{1cm} & \mbox{bz2} & \hspace{1.cm} & downstream z component of interplanetary magnetic field (nT); \\
		26\hspace{1cm} & \mbox{sat} & \hspace{1.cm} & satellite used for calculations: 1 for Wind, 2 for ACE \\
		27\hspace{1cm} & x & \hspace{1.cm} & x GSE position (in Re) of solar wind monitor at UTS; \\
		28\hspace{1cm} & y & \hspace{1.cm} & y GSE position (in Re) of solar wind monitor at UTS; \\
		29\hspace{1cm} & z & \hspace{1.cm} & z GSE position (in Re) of solar wind monitor at UTS; \\
		30\hspace{1cm} & \mbox{minSMR} & \hspace{1.cm} & minimum SMR within two hours of shock impact \\
		\hline
	\end{tabular}
	\caption{Names of the variables associated with each shock event in the list and shown in Listing \ref{listing2}. The numbers in the first column are the numbers of the fields in the list and shown in the header of the file \texttt{full\_shock\_list\_2023.txt}.} 
	\label{table1}
	\end{table}

	Below are the steps taken to include a specific solution for each shock in the \texttt{full\_shock\_list\_2023.txt} list. These are the major revisions made to the list in comparison to its previous versions:

	\begin{enumerate}
		\item A filter is passed on the data to replace bad data points (e.g., 1E+31) by nan values which are then replaced by interpolated/averaged values to a common time cadence for both data sets (IMF and solar wind parameter data).

		\item The satellite (Wind or ACE) must be in the solar wind upstream of the Earth (x $>$ 14 $R_E$).

		\item If data of both satellites are simultaneously available, the data set with data of superior quality is used for computations.

		\item Events with either $M_A$ or $M_s$ (or both) smaller than one are generally discarded. Events with such conditions are only included in the list if they trigger significant geomagnetic activity, such as SMR variations of at least 15 nT.

		\item The solution obtained from equations (1-5) closest to the median value is selected to be included in the list. If the difference between the maximum and minimum values of \thxn{} is larger than 30$^\circ$, the solution chosen for the list will be the one that shows more agreement with ground geomagnetic response, represented by the SuperMAG indices (SMR, SMU, SML, SME), as shown in previous publications \citep[e.g., ][]{Wang2006a,Oliveira2015a,Rudd2019,Oliveira2021b}.

	\end{enumerate}
	
	The list version published by \cite{Oliveira2015a} had 461 events, whereas the list published by \cite{Oliveira2018b} had 547 events. This current data base has more events (603) and has a number of additional solar wind parameters and shock properties, which are shown in Table \ref{table1}. The list time span, January 1995 to May 2023, includes two entire solar cycles (SC23 and SC24), the end of declining phase of SC22, and the beginning of ascending phase of SC25. Therefore, this data base provides a solid number of events for future statistical studies given an appropriate availability of data sets to be investigated. \par

	Figure \ref{shock_hist} represents general statistical features of the 603 events in the shock data base. Panel a shows yearly shock number distributions and Carrington-rotation of 25.38 days \citep{Carrington1863} averaged sunspot numbers (solid black line). This figure shows a clear correlation between the number of events and sunspot numbers, a result that is already well known \citep{Kilpua2015a,Oliveira2015a,Rudd2019}. Observations show that SC23 was significantly stronger than SC24 based on the overall sunspot numbers, which is reflected on the total number of shocks observed in the corresponding periods (432 and 187, respectively). \cite{Zhu2022b} predicted that SC25 will be stronger than SC24 and will reach its maximum value around July 2025. Therefore, according to these predictions, it is reasonable to expect that SC25 will have a similar number of shocks with respect to SC24. \par 

	The overall statistical properties of sunspot observations and shock parameters are quantified and shown in Table \ref{table2}. These results, including mean, median, percentile values, and shock number distribution correlation with sunspot numbers, are in excellent agreement with past studies \citep{Oliveira2015a,Kilpua2015a,Oliveira2018b,Rudd2019}.

	\begin{table}
		\centering
		\begin{tabular}{l c c c c c c}
			\hline
				\multicolumn{7}{c}{Shock parameters} \\
			\hline
				variable  &  LPT  &  median  &  mean  &  UPT  &  STD  & $\#$ of events \\
			\hline
				\thxn{} [$^\circ$] 	& 132.96   &  148.43  &  147.99  &  163.56  &   15.90 	&  603 \\
				\phiyn{} [$^\circ$] & -105.87  &   13.10  &    7.12  &  120.00  &  108.41 	&  603 \\
				\thbn{} [$^\circ$]  & 41.05    &   64.14  &   59.87  &   79.77  &   21.16 	&  603 \\ 
				\vs{} [km/s]		& 335.83   &  435.68  &  463.20  &  575.75  &  161.88	&  603 \\
				\Ma{}				& 1.74     &    2.58  &    3.13  &    4.05  &    1.95 	&  603 \\
   				\Ms{} 				& 1.33     &    1.86  &    2.16  &    2.79  &    1.18 	&  603 \\
   				\Xn{}				& 1.54     &    1.96  &    2.13  &    2.62  &    0.75 	&  603 \\
				\Xdp{} 				& 1.45     &    1.79  &    1.93  &    2.36  &    0.69	&  603 \\
   				\Xb{} 				& 1.81     &    2.49  &    2.98  &    3.86  &    1.56 	&  603 \\
			\hline
				\multicolumn{7}{c}{Sunspot numbers (complete solar cycles)} \\
			\hline
				solar cyle $\#$  &  LPT  &  median  &  mean  &  UPT  &  STD  & $\#$ of events \\
			\hline
				SC23 				&  13.00  	 &   63.00  &   82.39  &  150.00  &   72.74 &  342 \\
				SC24 				&   11.00    &   44.00  &   54.08  &   98.00  &   47.01 &  187 \\
		\end{tabular}
		\caption{Upper part: Shock parameters calculated from the shock data base released with this publication. The statistical data are: LPT, lower percentile (20\%); median value; mean value; UPT, upper percentile (80\%); and standard deviation (STD). There are 603 events in this shock list. Lower part: statistical results of sunspot number observations of the two complete solar cycles (SC) covered in the shock data base: SC23 and SC24.}
		\label{table2}
	\end{table}

	\setcounter{figure}{1}
	\renewcommand{\figurename}{Figure}
	\renewcommand{\thefigure}{ \arabic{figure}}

		\begin{figure}
			\centering
			\includegraphics[width = 18cm]{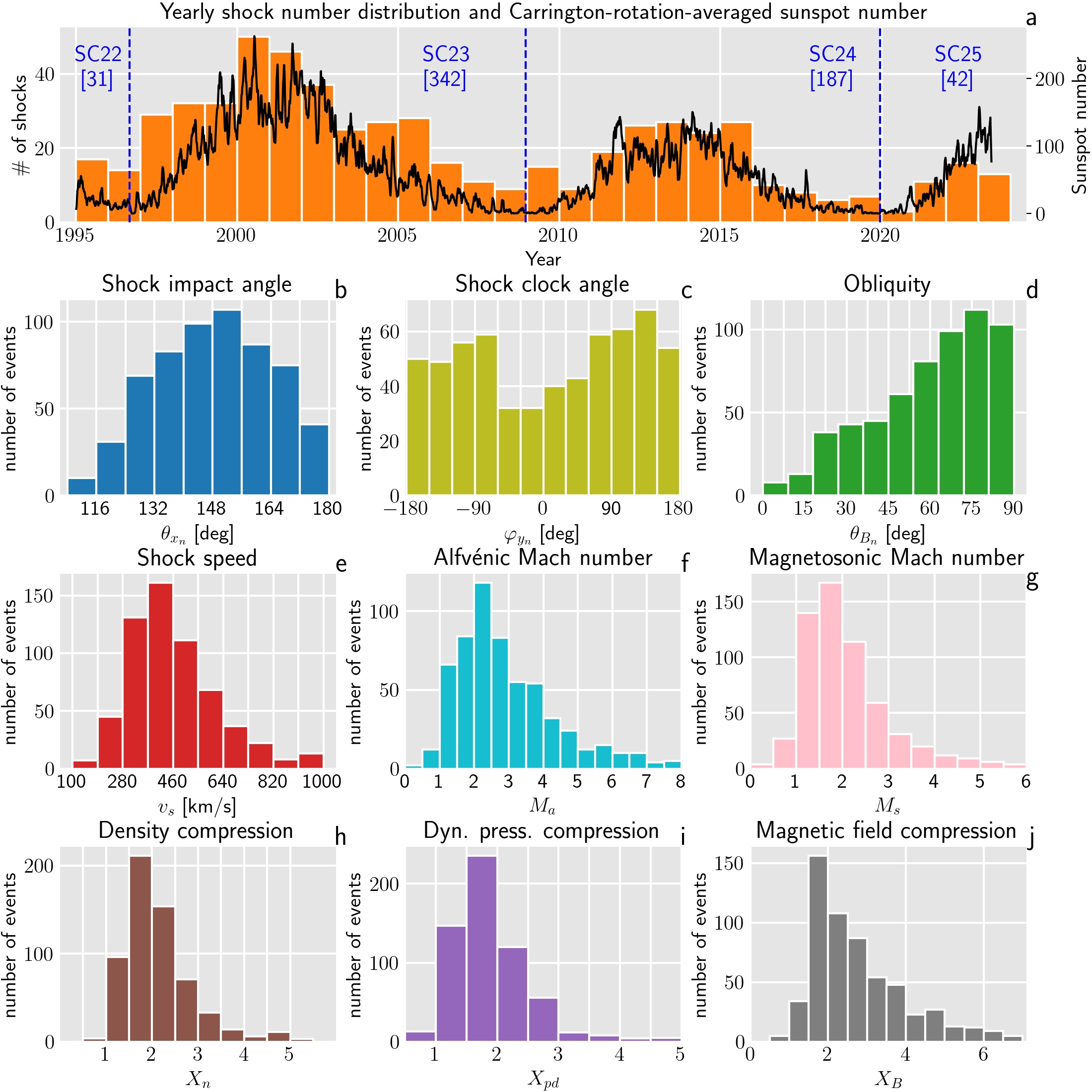}
			\caption{Statistical properties of shocks in the data base from January 1995 to May 2023 (603 shocks). Panel a: shock number distribution and Carrington rotation-averaged sunspot numbers. Panels b-j: number distributions of shock parameters obtained from equations 6-17.}
			\label{shock_hist}
		\end{figure}

\section{Suggested use of the IP shock data base}

	The IP shock data base described in this report can be used in many ways. For example, this list can be used in studies involving geomagnetic activity following shock impacts from the geospace to the ground, including particle acceleration by shocks, particle dynamics and energization in the radiation belts, magnetospheric ultra-low frequency (ULF) waves, magnetic field response in geosynchronous orbit, field-aligned currents, role of shocks in substorm triggering, ionospheric irregularities, high-latitude thermosphere response (neutral density and nitric oxide) to shock impacts, ground magnetometer response (dB/dt variations) and subsequent effects on GICs, and many others. Therefore, this shock list can be used in a variety of space physics and space weather investigations. \par

	A major feature of this shock list is the possibility of using the shock impact angle as a factor controlling geomagnetic activity. \cite{Oliveira2023b} has recently reviewed the effects of shock impact angles on the subsequent geomagnetic activity and also suggested a few topics for future research.

	\begin{enumerate}
		\item Shock impact angle effects on intensities and latitudinal extensions of \dbdt{} variations linked to enhancements of GICs \citep{Carter2015,Oliveira2018b,Oliveira2021b}.

		\item Role of shock inclinations in controlling the triggering and wave modes of ULF waves and their interaction with magnetospheric cold plasma and wave-particle interactions \citep{Oliveira2020d,Hartinger2022}.

		\item Effects caused by different shock orientations on thermospheric neutral and nitric oxide molecules that control thermosphere heating and cooling affecting the subsequent satellite orbital drag in low-Earth orbit \citep{Oliveira2019b,Zesta2019a}.

		\item Shock impact angle effects on the dynamics of radiation belts (e.g., particle acceleration, enhancements, dropouts, and loss of relativistic electrons in the magnetosphere) \citep{Tsurutani2016,Hajra2018b}.

		\item Role of shock impact angle in triggering magnetospheric super substorms, with minimum SML $<$ --2500 nT \citep{Hajra2018a,Tsurutani2023}
	\end{enumerate}

	Finally, I would like to urge researchers to perform numerical simulations of shocks with different orientations. For example, \cite{Welling2021} argued that the ``most perfect" CME would be very fast and impact Earth head-on. They performed numerical simulations of the impact of a perfect CME on the magnetosphere and concluded that ground \dbdt{} variations were noted in very low latitude regions because the CME impact was purely frontal. Furthermore, our shock list can be very useful in simulations comparing real observations with results yielded by numerical simulations of shocks with different orientations for many different space weather purposes.

\section*{Data Availability Statement}

	The IP shock data base can be downloaded from Zenodo (\url{https://zenodo.org/record/7991430}). The solar wind plasma and IMF data observed by Wind and ACE used to calculate the shock impact angles including shock properties were obtained from the CDAWeb (Coordinated Data Analysis) website provided by NASA Goddard Space Flight Center's Space Physics Data Facility (\url{http://cdaweb.gsfc.nasa.gov}). The geomagnetic index data used in this publication were downloaded from the SuperMAG initiative website: \url{https://supermag.jhuapl.edu}. Sunspot number data was downloaded from the SILSO (Sunspot Index and Long-term Solar Observations) website (\url{https://www.sidc.be/silso/datafiles}).

\section*{Conflict of Interest Statement}

	The author declares that the research was conducted in the absence of any commercial or financial relationships that could be construed as a potential conflict of interest.

\section*{Author Contributions}

	This data report article was written by the author without any direct contributions from others.

\section*{Funding}
	
	This work was possible thanks to the financial support provided by the NASA HGIO program through grant 80NSSC22K0756.

\bibliographystyle{frontiersinSCNS_ENG_HUMS} 
\bibliography{/Users/dennyoliveira/Documents/Papers/Notes/Oliveira_main}

\end{document}